\documentclass[aps,superscriptaddress,showpacs,twocolumn]{revtex4}
\usepackage{tabularx}
\usepackage{bm}
\usepackage{subfigure}
\usepackage{euscript}
\usepackage{epsfig,psfrag,subfigure}
\usepackage{graphicx,psfrag,subfigure}
\usepackage{color}
\usepackage{amsmath,amsfonts}
\usepackage{exscale}
\usepackage{amsbsy}
\usepackage{subfigure}
\usepackage{units}
\usepackage{epstopdf}

\def\be{\begin{equation}}
\def\ee{\end{equation}}
\def\ba{\begin{eqnarray}}
\def\ea{\end{eqnarray}}

\begin{document}

\preprint{INT-PUB-09-024}
\title{Theory of deflagration in disordered media}
\author{Mauro Schiulaz}
\affiliation{Department of Physics, University of Washington, Seattle, WA 98195, USA}
\author{Christopher R. Laumann}
\affiliation{Department of Physics, University of Washington, Seattle, WA 98195, USA}
\affiliation{Department of Physics, Boston University, Boston, MA, 02215, USA}

\author{Alexander V. Balatsky}
\affiliation{Institute for Materials Science, Los Alamos, NM 87545,  USA}
\affiliation{Nordita, Center for Quantum Materials, KTH Royal Institute of Technology and Stockholm University, Roslagstullsbacken 23, SE-106 91 Stockholm, Sweden}
\author{Boris Z. Spivak}
\affiliation{Department of Physics, University of Washington, Seattle, WA 98195, USA}

\begin{abstract}
The conventional theory of burning works well in the case of uniform media where all system parameters are spatially independent.
We develop a theory of burning in disordered media.
In this case, rare regions (hot spots) where the burning process is more effective than on average may control the heat propagation in an explosive sample.
We show that most predictions of the theory of burning are quite different from the conventional case.
In particular, we show that a system of randomly distributed hot spots exhibits a dynamic phase transition, which is similar to the
percolation transition. Depending on the parameters of the system the phase transition  can be either first or second order.
These two regimes are separated by a tricritical point.  The above results may be applicable to dynamics of any overheated disordered system with a first order phase transition.
\end{abstract}

\date{\today}
\maketitle

\section{Introduction}
Burning processes are related to exothermic reactions whose rates  are quickly increasing functions of the temperature.
Usually two regimes are considered:
``detonation,'' which propagates supersonically via shock waves,
and ``deflagration'' (``combustion''), which propagates subsonically as determined by heat diffusion.
In the deflagration regime, the heat transferred by diffusion processes can be described by a nonlinear heat conduction equation~\cite{Landau:1987ab,dremin1999toward,lee2008the,Frank-Khamenetski:1955aa,Zeldovich:1980aa}
\begin{equation}\label{heatcondEq}
(\partial_{t}-\frac{\kappa}{C} \partial_{{\bf r}}^{2})T({\bf r},t)=\frac{p_{0}}{C}e^{-\frac{U}{T}}-\frac{T-T_{0}}{\tau},
\end{equation}
where $C$ and $\kappa$ are the (volumetric) specific heat and the thermal conductivity, respectively.  $T_0$ is the temperature of the environment and $\tau$ is a temperature relaxation time.
The parameters $p_0$ and $U$ describe the heat produced and activation energy of the exothermic  reaction.

The conventional theory of deflagration describes systems where all parameters in Eq.~(\ref{heatcondEq}) are spatially homogeneous.
In this case, the width of the flame front and its velocity are of order \cite{Landau:1987ab}
\begin{equation}
\lambda_{\textrm{front}}\sim \sqrt{\kappa \tau_r} , \qquad V_{\textrm{front}}\sim \sqrt{\frac{\kappa}{\tau_r}}
\end{equation}
respectively.
Here, $\tau_r$ is a characteristic reaction (depletion) time for the exothermic reaction.

Generally, in solid explosives all parameters in Eq.~\eqref{heatcondEq} are random, sample specific functions of coordinates.
Moreover, the ignition of exothermic chemical reactions in disordered solid explosives has long been known to occur after locally heated regions of the material, called ``hot spots'', are formed by various processes
(see, e.g., Refs.~\cite{bowden1985initiation,Field:1992aa,Tarver:1996aa,McGrane:2009aa}).
In many cases, these are thought to control heat propagation.
The existence of hot spots, where the burning process is more effective than on average, can be related to the tails of the distribution function of these parameters.
We show below that in this case most of the predictions of the theory of burning are quite different from the conventional case \cite{Landau:1987ab,dremin1999toward,lee2008the,Frank-Khamenetski:1955aa}.
In particular, we show that a system of randomly distributed hot spots exhibits a dynamic phase transition, which in a certain regime is similar to the percolation transition.
Once started, an explosion either is able to propagate through the entire sample, or it stops after burning only a finite fraction of the system. Depending on parameters of the system the phase transition can be either first or second order, which are separated by a tricritical point.

To illustrate the origin of the hot spots in the deflagration regime, let us first consider an explosive of radius $r_0$ embedded into a uniform neutral region where $p_{0}=0$ and the heat dissipation is 
negligible ($\tau \to \infty$) \cite[p. 199]{Landau:1987ab}.
Using the conventional approximation $1/T\approx 1/T_{0}-(T-T_{0})/T_{0}^{2}$, the right hand side of Eq.~(\ref{heatcondEq}) becomes
\begin{equation}
p_{0}\exp(-U/T)\sim p_{0}({\bf r})\exp(\theta(T-T_{0})),
\end{equation}
where $\theta=U/T_{0}^{2}$, and $p_0\sim \rho$ is proportional to the density of explosive material, $\rho$, inside the hot spot.

In order to find stable solutions for the hot spot, the heat released per unit time%
, \mbox{$\sim p_0 r_0^3(1 + \theta(T - T_0))$},
must balance the heat transferred to the neutral region%
, \mbox{$\sim \kappa r_0^2 (T - T_0)$}%
.
These produce a stable solution if
\begin{equation}
\frac{p_{0}\theta r_{0}^{2}}{\kappa} < 1,
\end{equation}
in which case $T_{\textrm{center}}-T_{0}\sim 1/\theta$.
Here $T_{\textrm{center}}$ is the temperature in the center of the hot spot.
A more detailed treatment provides an expression for the critical temperature $T_{c}$ of explosion of the hot spot (see Ref.~\cite{Landau:1987ab}),
\begin{equation}
\label{Tc}
T_{c}\sim \frac{U}{\ln ((p_{0}\theta r_{0}^{2})/\kappa) }
\end{equation}
Generally, Eq.~(\ref{heatcondEq}) should be supplemented with an equation describing the dynamics of the density of the explosive, $\rho(t)\sim p(t)$.
If the temperature near a hot spot exceeds the critical one, an  explosion begins, $\rho(t)$ decreases, and eventually the whole hot spot is burned.
As a result, an energy $Q_{i}$ is released.
This propagates through the neutral medium to another hot spot and ignites it.
If the hot spots are distributed sufficiently far from each other, we can neglect the burning time of an individual hot spot compared to the time of intersite temperature propagation.
In this approximation Eq.~(\ref{heatcondEq}) reduces to the following:
 \begin{equation}\label{heatEq1}
(\partial_{t}-\frac{\kappa}{C} \partial_{{\bf r}}^{2})T({\bf r},t)=\sum_{i}\frac{Q_{i}}{C}\delta({\bf r}-{\bf r}_{i})\delta(t-t_{i})-\frac{T-T_{0}}{\tau}.
\end{equation}
Here $t_{i}$ is the time at which the temperature at the $i$th hot spot reaches its critical value $T({\bf r}={\bf r}_{i})=T_{c}^{(i)}$, and ${\bf r}_{i}$ is the position of the spot.
Generally, both $T_{c}^{(i)}$ and $Q_{i}$ are random quantities.
Stability at initial temperature $T_0$ requires that the distribution function of the critical temperatures, $P(T_{c})$, vanishes for $T_{c}<T_0$.

Let us investigate whether an explosion, once started, is able to propagate through the entire sample, or stops after consuming only a finite fraction of the system.
To illustrate the situation we start with the relatively simple case of large dissipation (small $\tau$) and assume that the values of $Q_{i}=Q$ are the same for all hot spots.
In Secs.~\ref{sec:numerics}~and~\ref{sec:asymmetric_percolation_model}, wel consider a more general situation where both $T_{c}^{(i)}$ and $Q_{i}$ are randomly distributed, and find that the situation does not change qualitatively.

The heat released from the explosion of a single hot spot at the origin propagates in a diffusive temperature wave,
\begin{align}
	T(\mathbf{r}, t) &= \frac{Q/C}{(4\pi D t)^{d/2}} \exp\left(-\frac{r^2}{4Dt} - \frac{t}{\tau} \right),
\end{align}
where the diffusion constant $D = \kappa/C$, and $r$ is the distance from the origin.
This wave ignites hot spots at position $\mathbf{r}_i$ if the local temperature rises above the critical temperature $T_{ic}$.
The peak of the wave arrives at distance $r$ at time
\begin{align}
	t^* = \frac{d \tau}{4}(\sqrt{\frac{4}{d^2} \frac{r^2}{l^2} + 1} - 1),
\end{align}
where $l = \sqrt{D\tau}$ is the dissipation length of the system.

\begin{figure}[ptb]
\includegraphics[scale=0.15]{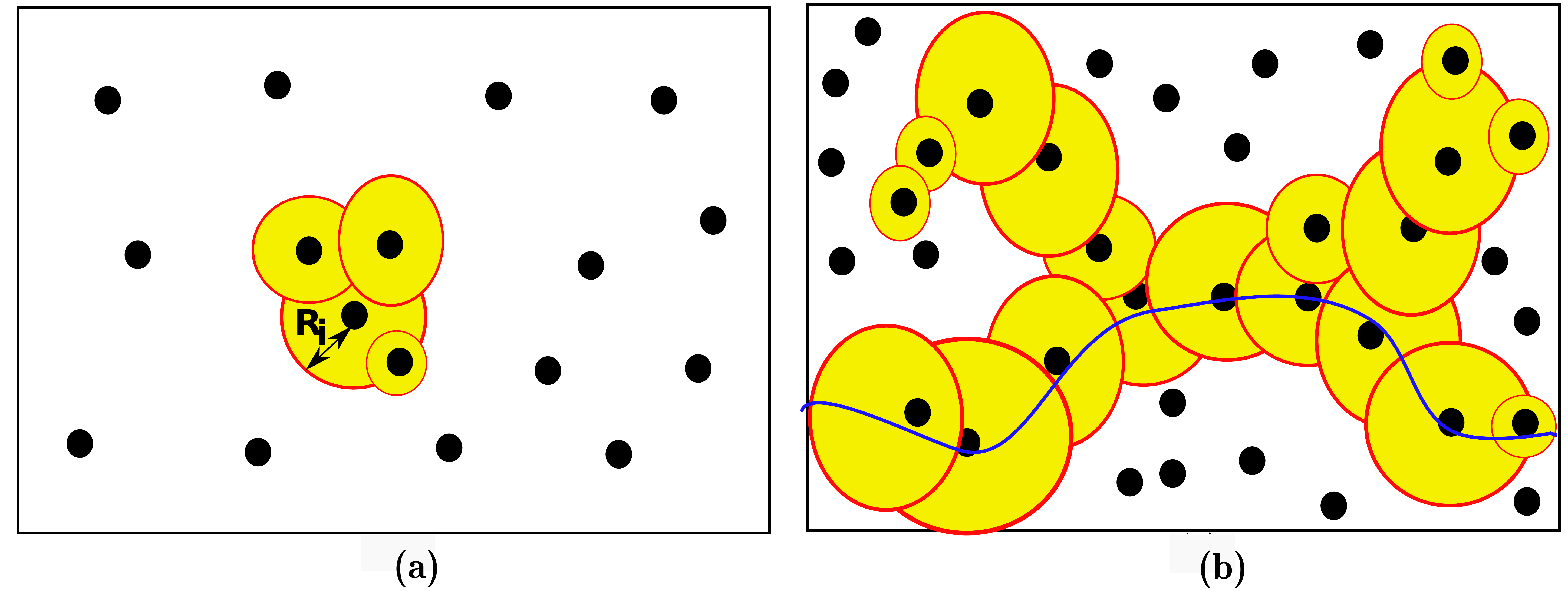}
\caption{
Pictorial representation of the mapping of the explosion process to percolation in the strongly dissipative regime.
A two/dimensional sample is composed by an inert material (white background), in which small explosive hot spots are embedded (black circles).
The explosion of any hot spot within a bubble of radius $R_i$ [see Eq.~(\ref{eq:bubbleradius})] around the $i$th hot spot will cause the $i$th spot to ignite as well.
(a) For $Q<Q_c$, a finite number of sites is activated.
(b) For $Q>Q_c$, an infinite cluster of exploded sites exists.
}
\label{fig:percolation}
\end{figure}

In the strongly dissipative regime, where the typical separation between hot spots $r \gg l$, the peak arrives at time
\begin{align}
	t^* \approx \frac{1}{2} \frac{\tau}{l} r,
\end{align}
producing an exponentially decaying maximum temperature,
\begin{align}
	T^*(r) \approx \frac{Q/C}{(2\pi l r)^{d/2}} \exp\left({-\frac{r}{l}}\right).
	\label{eq:T*}
\end{align}
In this regime, we can ignore the accumulation of heat from multiple diffusion waves: the residual heat from earlier explosions (more than one $t^*$ in the past) is an exponentially small correction to the wave due to the most recent explosions.
Thus, to each hot spot $i$ we can associate a bubble of radius $R_i$ satisfying
\begin{align}
\label{eq:bubbleradius}
	T^*(R_i, Q) &= T_{ic}.
\end{align}
Any hot spot which explodes within this bubble ignites the $i$th spot.
This maps the deflagration of the sample onto the percolation problem defined by the bubbles; see Fig.~\ref{fig:percolation}.
Thus, we expect there to be a $Q_c$ such that for $Q > Q_c$, the explosion percolates through the sample, while for $Q < Q_c$, the explosion propagates only up to the correlation radius
\begin{align}
	\xi &\sim |Q - Q_c|^{-\nu},
\end{align}
where $\nu$ is the correlation length exponent for percolation.
For example, $\nu = 4/3$ in the two-dimensional case.
The critical value of $Q_c$ can be estimated by $R_* \sim a$ where
\begin{align}
	a = \left(\int_{T_0}^\infty P(T_c) dT_c\right)^{-1/d}
\end{align}
is the typical separation between hot spots and $R_*$ is the typical radius of the bubbles [defined by Eq.~\eqref{eq:bubbleradius}].

In this regime, we are led to a picture of the explosion process which is completely different from the conventional one \cite{Landau:1987ab}.
At $Q>Q_{c}$ the explosion propagates along the percolating cluster, and the burning front width
\begin{equation}
\lambda_{\textrm{front}}\sim \xi
\end{equation}
diverges at the transition point, while the front velocity
\begin{equation}
 V_{\textrm{front}}\sim \frac{D}{a}
\end{equation}
is determined by the rate of heat diffusion along the percolation cluster. 
In Sec.~\ref{sec:numerics}, we present numerical simulations of Eq.~(\ref{heatEq1}) in two spatial dimensions, which substantiate the mapping to percolation at large dissipation (small $\tau$).
In the opposite, weakly dissipative, limit where the typical separation between hot spots $r \ll l$, the peak temperature arrives diffusively at
\begin{align}
	t^* &\approx \frac{1}{2d} \frac{r^2}{D},
\end{align}
carrying with it the volume averaged heat,
\begin{align}
\label{eq:tmaxnodiss}
	T^* &\approx \frac{Qe^{-1/2d} }{C(2\pi/d)^{d/2} } \frac{1}{r^d}
\end{align}
In this regime, heat accumulation is non-negligible and the deflagration process cannot be mapped onto static percolation.
Rather, if a small number $n$ of hot spots are close enough to ignite one another, even though the bubble network defined by Eq.~\eqref{eq:bubbleradius} is below percolation, the accumulated heat released, $n Q$, may ``jump'' gaps in the network.
The dynamics in this regime are driven by the disorder fluctuations at short distances and, as we see numerically, the system exhibits a dynamical first order phase transition in which the average size of the exploded cluster jumps to infinity discontinuously. This situation is somewhat reminiscent of ``bootstrap percolation'' models 
~\cite{ADLER1991453},
which are known to display either a first or second order transition, depending on the parameters and dimensionality of the model.

\begin{figure}[ptb]
\includegraphics[scale=0.7]{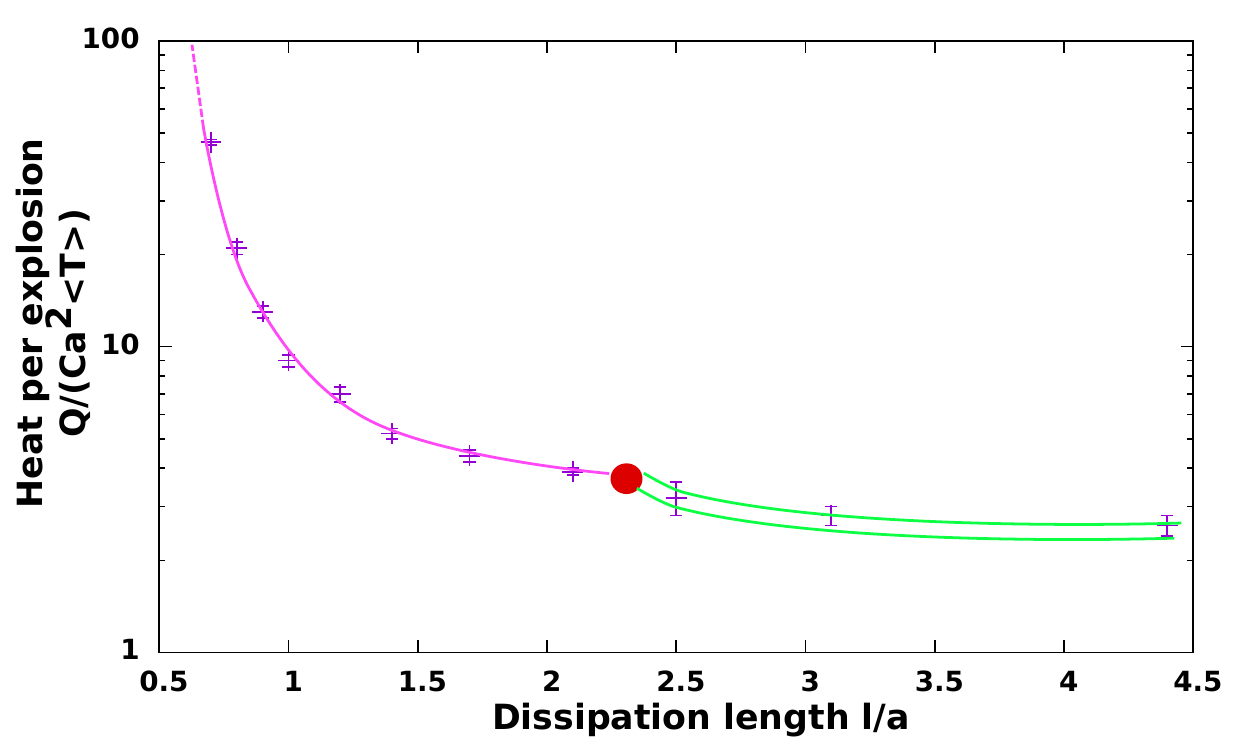}
\caption{Phase diagram of the uniform $Q$ model as a function of the emitted heat per explosion $Q$ and the dissipation length $l$.
The  transition line is first order for large $l$ (green double line) and second order for small $l$ (purple single line). The lines are meant to be a guide to the eye. The two regimes are separated by a multicritical point (red dot).}
\label{fig:phasediagram}
\end{figure}

The two regimes (the first and the second order phase transitions) are separated by a multicritical point at $l=l_c$.
We estimate the critical relaxation time as a time of diffusion on an average distance between the hot spots $\tau_{c}\sim \nicefrac{a^2}{D}$ or, in terms of lengths, when $l \sim a$.
The resulting phase diagram is shown in Fig.~\ref{fig:phasediagram}, where $l$ is measured in units of $a$, and $Q$ is measured in units of $C a^2 <T>$, where $<T>$ is the average value for the hot spots' critical temperature [see Eq.~(\ref{eq:PTc})].

\section{Numerical analysis of the dynamical explosion model}
\label{sec:numerics}

We assume each sample to be formed by point like hot spots, placed on a two-dimensional square lattice with lattice spacing $a$ (which is taken as the reference length, in units of which distances are measured).
We also take periodic boundary conditions.
Each hot spot is assigned a critical temperature $T_{c}^{(i)}$ drawn from the distribution given by Eq.~\eqref{eq:PTc}.
For each sample we initialize the explosion by making a single randomly chosen site explode, and release an energy $Q_{i}$.
We set to 1 the specific heat $C$ of our sample, and therefore measure heat and temperature in the same units.
For the distribution of the critical temperatures, we take
\begin{equation}
P\left(T_{c}\right)\propto\begin{cases}
\left(T_{c}-T_{0}\right)^{\alpha}(T_{\textrm{max}}-T_c)^{\alpha}, & T_{0}\leq T_{c}\leq T_{\textrm{max}}\\
0, & \textrm{otherwise}.
\end{cases}
\label{eq:PTc}
\end{equation}
The average value for the critical temperature is $<T>=(T_{\textrm{max}}-T_0)/2$. In the simulations, in Eq.~(\ref{eq:PTc}) we have set $T_0=0$, $T_{\textrm{max}}=10$ and we have chosen the exponent $\alpha=4$.
\footnote{ From a physics consideration the value of $\alpha>0$ should be positive in order to ensure that our system is stable against thermal fluctuations. Indeed, for $\alpha=0$, an infinitely small increase of temperature activates a fraction of sites which scales  as $\log(L/a)$, where $L$ is the sample size.}

We simulate Eq.~(\ref{heatEq1}) discretizing time in steps $\Delta t$. For each sample, we initialize the explosion by making a single randomly chosen site explode, and release an energy $Q$. At each time step we let the heat propagate according to Eq.~(\ref{heatEq1}). Additionally, whenever  at site $i$ the temperature is higher than the local critical temperature, i.e. if $T({\bf r}_i,t)\geq T_c^{(i)}$, site $i$ explodes, releasing an extra energy $Q$, and then becomes exhausted (and cannot explode any longer).  In practice, this is implemented by repeating the following procedure at each time step $\Delta t$, for each site $i$:

\begin{itemize}

\item If $\ensuremath{T({\bf r}_{i},t)\geq T_{c}^{(i)}}$, the site explodes, releasing an extra energy $Q$:

\begin{equation}
T\left(\mathbf{r}_{i},t\right)\rightarrow T\left(\mathbf{r}_{i},t\right)+Q.
\end{equation}

 Each site can explode only once: after the detonation, it is considered as exhausted, and cannot explode any longer.

\item To represent heat diffusion, the temperature is then set as the average temperature of the surrounding environment:

\begin{equation}
T\left(\mathbf{r}_{i},t+\Delta t\right)=\frac{T\left(\mathbf{r}_{i},t\right)+\sum_{j=1}^{Z}T\left(\mathbf{r}_{j},t\right)}{Z+1},
\end{equation}

where $Z$ is the coordination number of the lattice, and $j$ labels the nearest neighbors of site $i$. This request fixes the value of $\Delta t$ as

\be
\Delta t=(Z+1)^{\frac{2}{d}}\frac{a^2}{4\pi D}.
\ee

This procedure is correct only asymptotically, since at short times it distorts the dynamics. However, this choice allows us to greatly increase the efficiency of the simulations, and we do not expect it to have any significant impact on the physics of the system. Its only possible effect is to introduce a systematic deviation in the {\em position} of the transition point, but not to change its nature.

\item Finally, the heat dissipation acts:

\begin{equation}
T\left(\mathbf{r}_{i},t+\Delta t\right)\rightarrow T_0+[T\left(\mathbf{r}_{i},t+\Delta t\right)-T_{0}]e^{-\frac{\Delta t}{\tau}}.
\end{equation}

\end{itemize}

\subsubsection{Numerical results at strong dissipation}\label{sec:strongdissipation}

\newcommand{\NN}{N_{\textrm{exp}}}

\begin{figure}[ptb]
\includegraphics[scale=0.7]{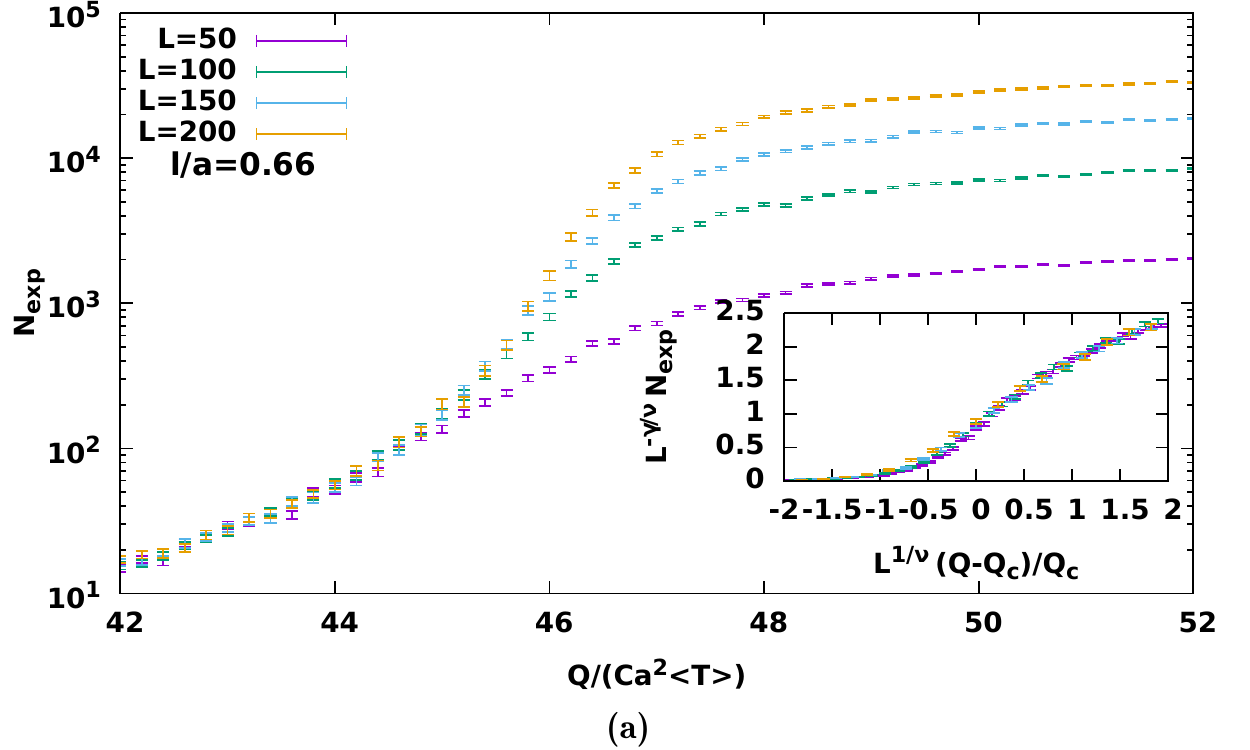}
\includegraphics[scale=0.7]{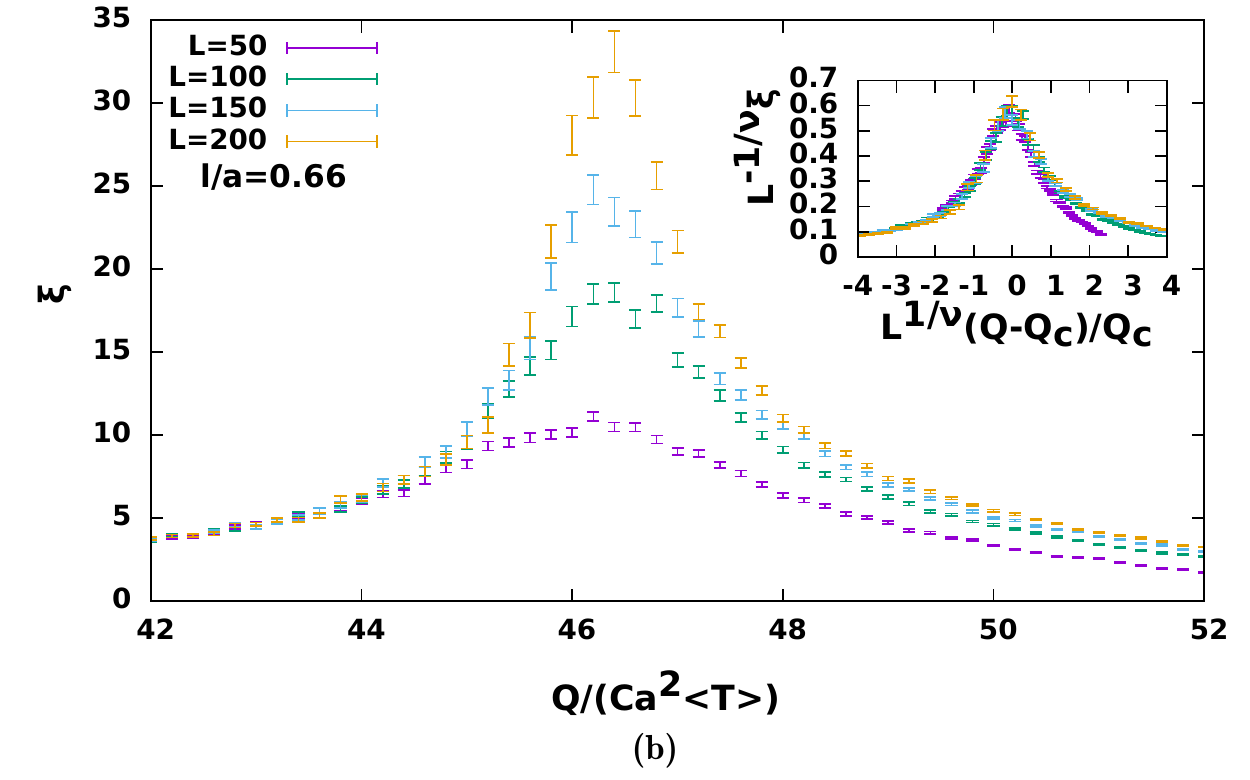}
\caption{%
Numerical results for the uniform $Q$ model, for $\nicefrac{l}{a}=0.66$ {\em (strong dissipation)}.
(a) The number of exploded sites, $\NN$, plotted as a function of $Q$, for systems of linear size $L=50a$ (purple line), $L=100a$ (green), $L=150a$ (blue), and $L=200a$ (orange). A transition from a regime in which $\NN$ is finite to one in which it is extensive is observed. Finite size scaling of the data, using the scaling ansatz of Eq.~(\ref{eq:scaling}) ({\em inset)}, is consistent with percolation.
(b) Correlation length $\xi$ as a function of $Q$. A divergence with system size at the transition point $Q_c/(Ca^2<T>)\approx46.4$ is observed, in accordance with the presence of a second order transition. Finite size scaling of the data ({\em inset}) shows that the critical exponent is compatible with $\nu=4/3$.}\label{fig:dynamicalstrong}
\end{figure}

 Figure~\ref{fig:dynamicalstrong}(a) shows the average over disorder realizations of the number of exploded sites $\NN$, which are initiated by one explosion  as a function of the heat released per explosion, $Q$, for $\nicefrac{l}{a}=0.66$. For $Q<Q_{c}$ the explosion involves only a finite number of hot spots, while for $Q>Q_c$, $\NN$ grows with system size.

In order to compare our results against percolation theory, we have performed a finite size scaling of the data, using the following scaling ansatz:
\begin{equation}
\NN(Q,L)\propto L^{\frac{\gamma}{\nu}}f\left(L^{\frac{1}{\nu}}\frac{Q-Q_{c}}{Q_{c}}\right),\label{eq:scaling}
\end{equation}
where $L$ is the linear size of the sample, $f$ is a universal function, and $\gamma$ and $\nu$ are scaling exponents. In particular, $\nu$ governs the divergence of the correlation length $\xi$ at the critical point, while $\gamma$ describes the divergence of the mean cluster size. We have then extracted the values of $Q_c,\nu$, and $\lambda$ that optimize the collapse of the curves onto each other. The collapse of the data is shown in the inset of Fig.~\ref{fig:dynamicalstrong}(a). This procedure yields $Q_c/(Ca^2<T>)\approx46.8\pm0.4$, $\nu\approx1.33\pm0.03$ and $\gamma\approx2.35\pm0.05$. These values are in agreement with the predictions of percolation theory, $\nu=\nicefrac{4}{3}$ and $\gamma=\nicefrac{43}{18}$~\cite{stauffer1994introduction,meester1996continuum}.

 The correlation length $\xi(Q)$ of a pattern of exploded hot spots has been computed in the standard way~\cite{stauffer1994introduction,meester1996continuum}: at $Q<Q_{c}$, where the explosion does not percolate, as shown in Fig.~\ref{fig:percolation}(a), the correlation length is defined as the mean distance between sites belonging to the exploded cluster,
\begin{equation}
	\xi\equiv\sqrt{\frac{2}{N_c(N_c-1)}\sum_{i,j\in c}(r_i-r_j)^2}.
	\label{eq:xi}
\end{equation}
In the above, $N_{c}$ is the total number of hot spots belonging to the exploded cluster $c$. Analogously, in two dimensions, for $Q>Q_{c}$, the correlation length can be defined as the mean distance between sites belonging to the same connected cluster of unexploded sites. The obtained data are plotted in Fig.~\ref{fig:dynamicalstrong}(b).

As before, to numerically study the divergence of the correlation length, we perform a finite size scaling of our data:
\begin{equation}
	\xi(Q,L)\propto L^{\nicefrac{1}{\nu}}g\left(L^{\frac{1}{\nu}}\frac{\left|Q-Q_{c}\right|}{Q_{c}}\right).
\end{equation}
The optimal collapse, shown in the inset of Fig.~\ref{fig:dynamicalstrong}(b), is found with $Q_c/(Ca^2<T>)\approx46.4\pm0.2$ and $\nu\approx1.34\pm0.4$, in agreement with the values extracted from $\NN$.

Thus we conclude that, for strong dissipation, our model belongs to the universality class of two-dimensional percolation theory.

\subsubsection{Numerical results at weak dissipation}\label{sec:weakdissipation}

\begin{figure}[ptb]
\includegraphics[scale=0.7]{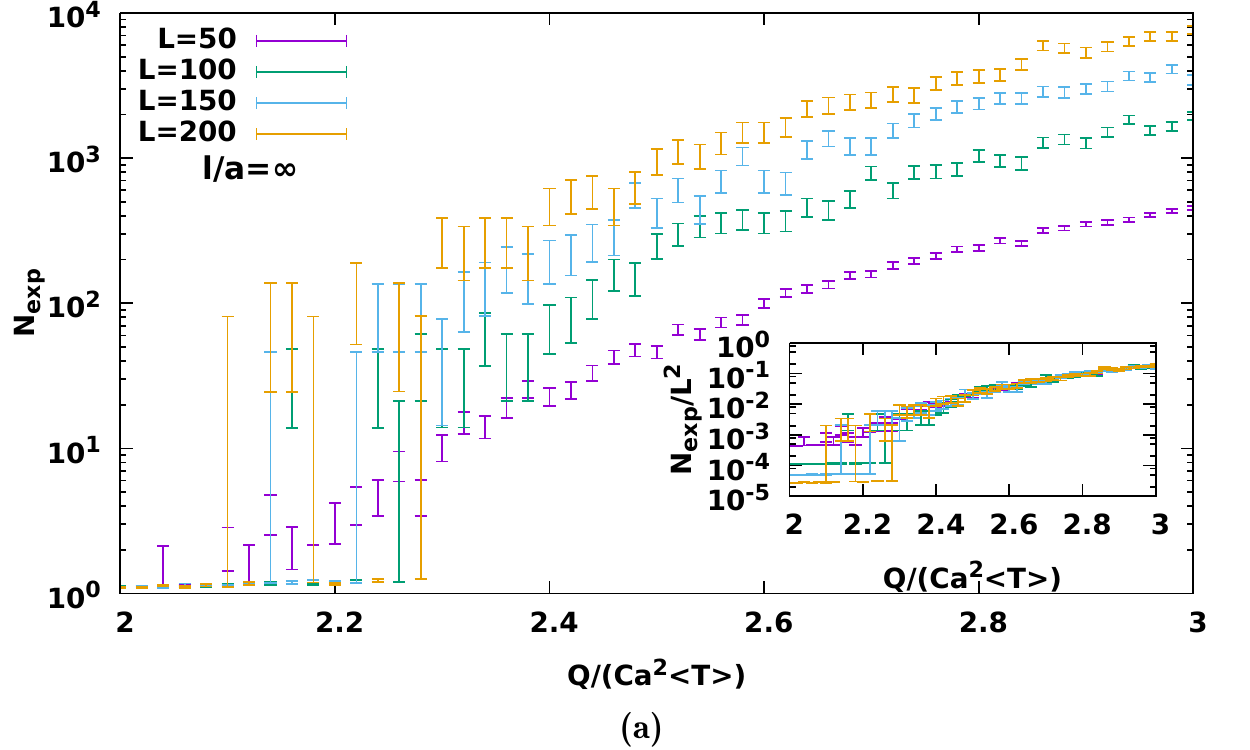}
\includegraphics[scale=0.7]{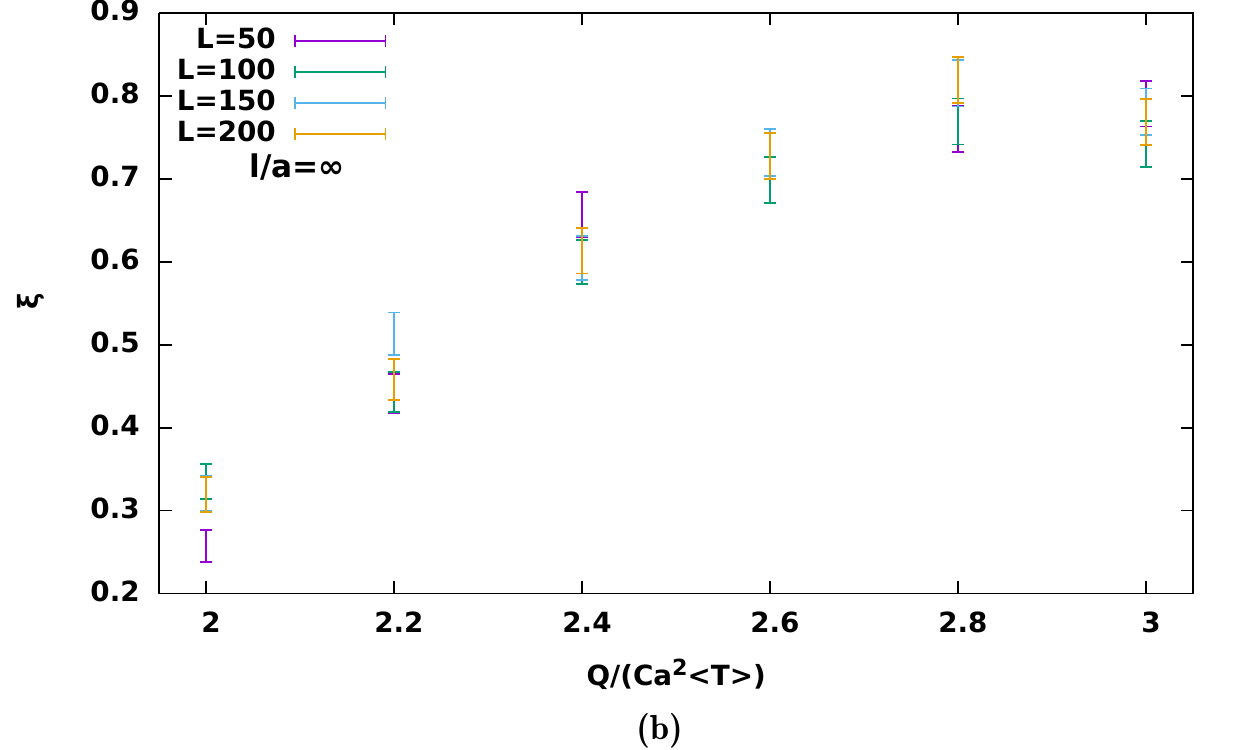}
\caption{
Numerical results for the uniform $Q$ model, for $l=\infty$ {\em (no dissipation)}. (a) The number of exploded sites, $\NN$, plotted as a function of $Q$. The transition is now first order, with a coexistence region characterized by very large fluctuations. The plot of the fraction of exploded volume, $n_{\textrm{exp}}$ {\em (inset)}, and direct computation of the correlation length (b) both show that no diverging length scale can be identified at the transition point: the transition is discontinuous.}
\label{fig:dynamicalweak}
\end{figure}

The results of numerical simulations at large values of $l$, presented in Fig.~\ref{fig:dynamicalweak}, are very different from the previous case (Fig.~\ref{fig:dynamicalstrong}).
From the main plot, it can be seen that $\NN$ no longer grows continuously as a function of $Q$, but rather jumps from being of $O(1)$ to being proportional to the total volume.

Additionally, no finite size scaling procedure is needed to make the curves at various system sizes collapse onto one another. Indeed, for small $Q$, the bare curves fit onto each other [Fig.~\ref{fig:dynamicalweak}(a)]; for large $Q$ one simply has to plot the fraction $n_{\textrm{exp}}(q)\equiv\nicefrac{\NN}{V}$, $V$ being the total volume, to get the desired collapse [inset of Fig.~\ref{fig:dynamicalweak}(a)]. These two regimes are separated by a ``coexistence'' region, characterized by very large fluctuations: in this region, the activation of a single site has finite probability of either triggering an explosion or doing nothing.

The direct computation of the correlation length, shown in Fig.~\ref{fig:dynamicalweak}(b) indicates that $\xi$ does not diverge, but rather is  independent from system size.
These are characteristic features of a first order phase transition. Such a first order transition line is separated from the second order one by a tricritical point, which we find at $\nicefrac{l_c}{a}\approx2.3\pm0.1$.

 \subsubsection{Numerical results with $Q_{i}$ and $T_i$  correlated}\label{sec:QproptoTc}

\begin{figure}[ptb]
\includegraphics[scale=0.7]{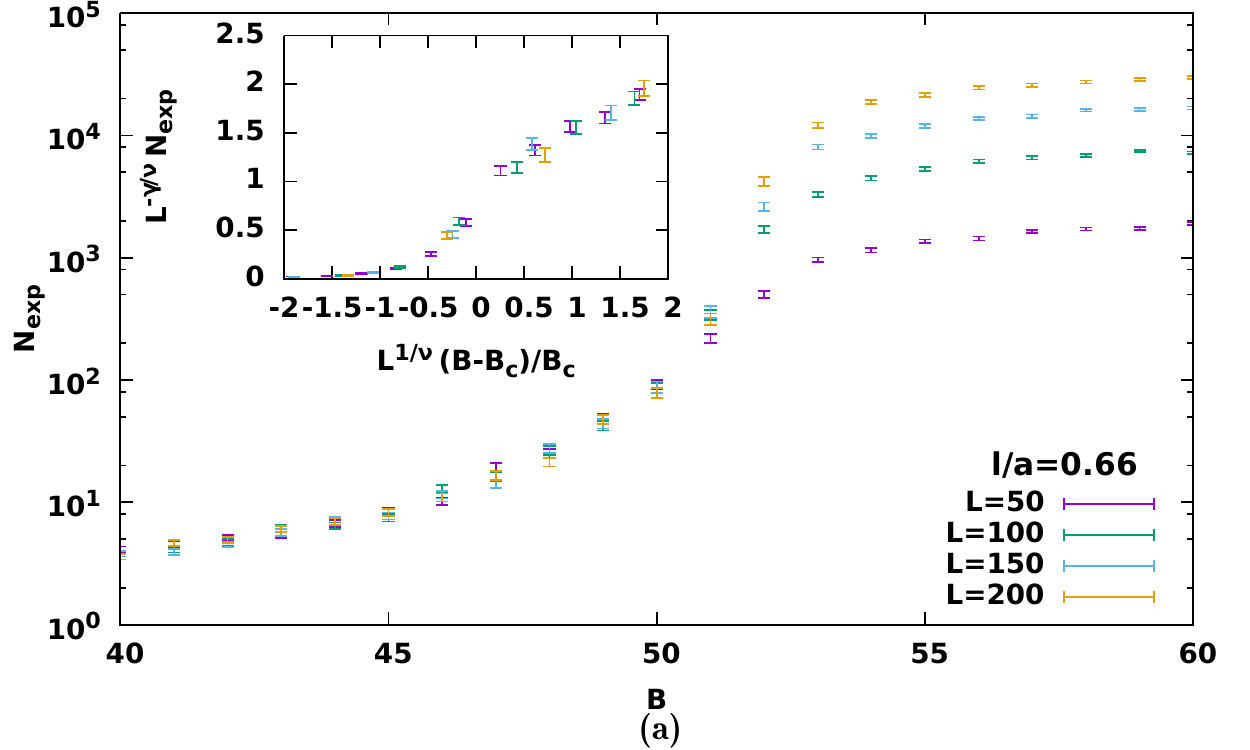}
\includegraphics[scale=0.7]{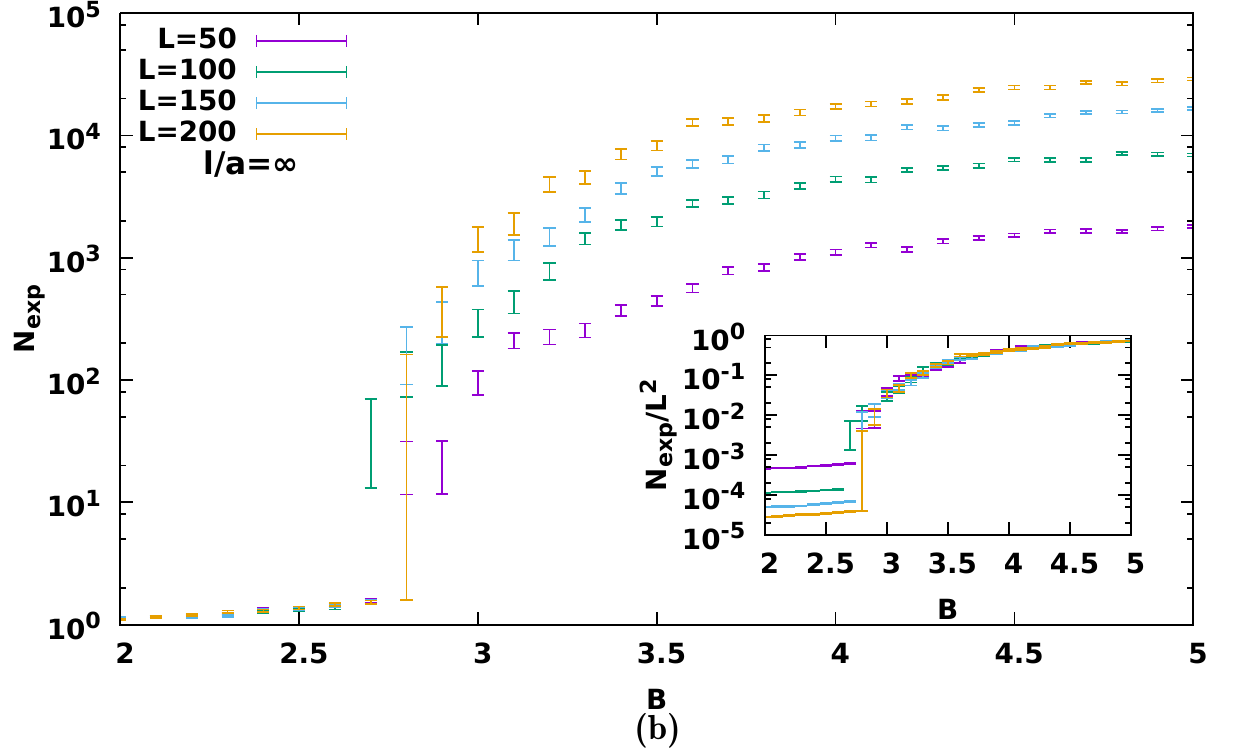}
\caption{Number of exploded sites, $\NN$, as a function of the proportionality constant $B$, for a system with $T_c$ proportional to $Q$. (a) $\nicefrac{l}{a}=0.66$ (strong dissipation); (b) $l=\infty$ (no dissipation). The transition is of second order in (a), while it is first order in (b). This is confirmed by the collapse of the data points by finite size scaling, shown in the insets.}
\label{fig:QproptoTc}
\end{figure}

In real systems, both $T_i$ and $Q_{i}$ are random quantities which may be  correlated.
We show below in this case the phase diagram shown in Fig.~\ref{fig:phasediagram} does not change qualitatively.
To illustrate this fact let us consider a simple case
\begin{equation}
Q_i=B C a^2 (T_c^i-T_0),
\label{eq:QproptoTc}
\end{equation}
where $B>0$ is a dimensionless proportionality constant.
The results of numerical simulations are presented in Fig.~\ref{fig:QproptoTc}, where  the average number of exploded sites, $\NN$, is shown as a function of the proportionality constant $B$. Figure~\ref{fig:QproptoTc}(a) plots the data for $\nicefrac{l}{a}=0.66$ (strong dissipation regime), while Fig.~\ref{fig:QproptoTc}(b) shows the dissipationless regime $l=\infty$. The explosion is initialized at a random hot spot.

The data have been analyzed using the same procedure explained in Secs.~\ref{sec:strongdissipation} and~\ref{sec:weakdissipation}, respectively. For $\nicefrac{l}{a}=0.66$, we found the transition point to be $B_c\approx52.3\pm0.1$, the exponent for the correlation length to be $\nu\approx1.33\pm0.05$, and the exponent for the average number of exploded sites to be $\gamma\approx2.3\pm0.1$. The transition between the two regimes takes place at a finite value $\nicefrac{l_c}{a}=1.5\pm0.1$, whose value is a factor $1.5$ smaller than its equivalent in the uniform $Q$ model.

\section{Asymmetric Percolation Model}
\label{sec:asymmetric_percolation_model}

The model described by Eq.~(\ref{heatEq1}) is dynamical.
In order to clarify its relation with percolation theory, we introduce a simplified static model.
If the heat released by the explosion of site $i$ is capable of igniting site $j$ [which means that $T^*(r_{ij})\geq T_c^j$; see Eq.~(\ref{eq:T*})], we draw a {\em directed} link from site $i$ to site $j$.

\begin{figure}[ptb]
\includegraphics[scale=0.3]{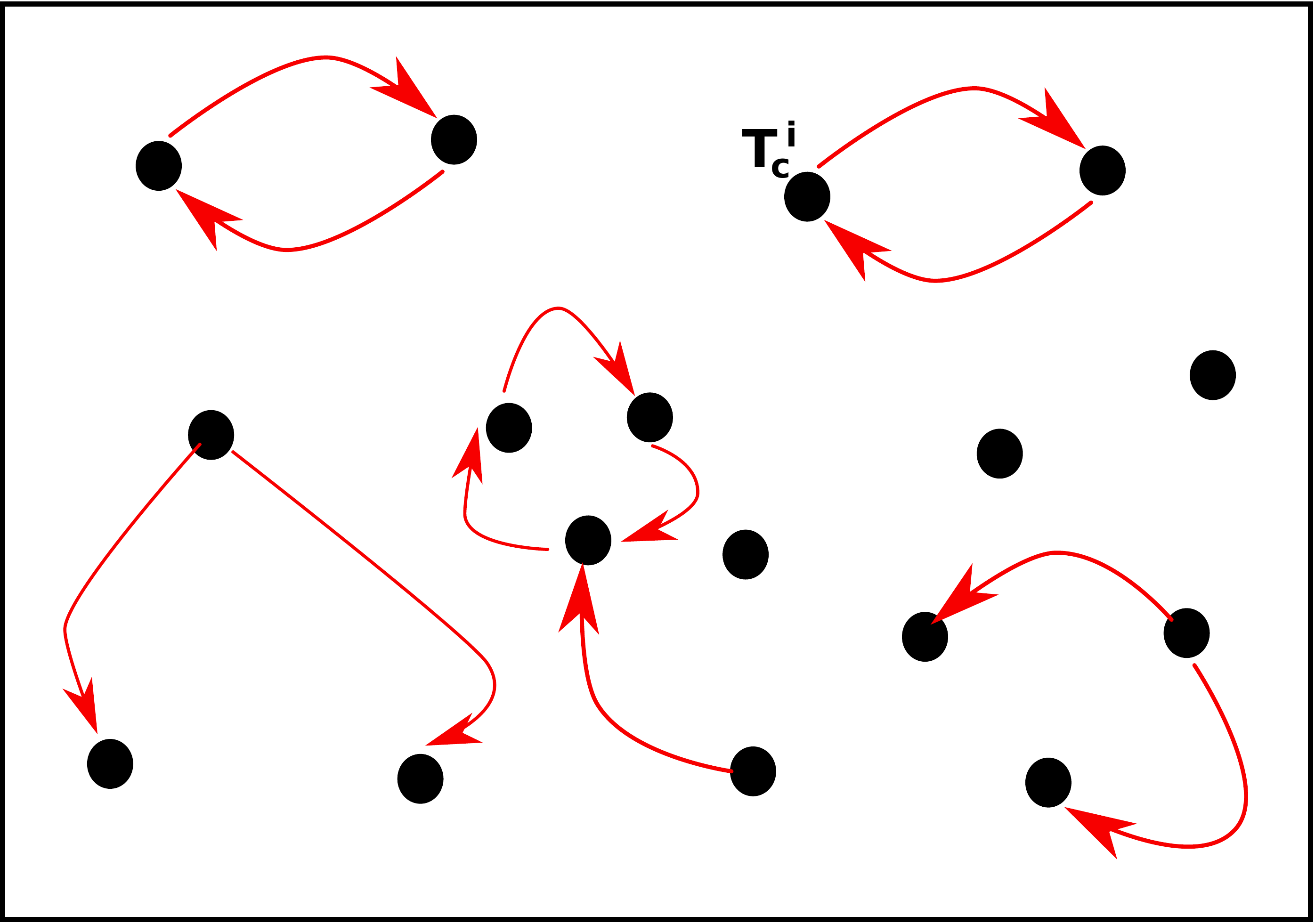}
\caption{Pictorial representation of the asymmetric percolation model.
Each site is assigned a critical temperature $T_c^i$. Then for each site $j$ a directed link is drawn from $i$ to $j$ if $T^*(r_{ij})\geq T_c^j$.}
\label{fig:asymmetricart}
\end{figure}

In general, this relation is asymmetric--that site $i$ is linked to site $j$ need not imply the opposite.
Thus, each pair of sites can be connected by zero, one, or two  directed links.
Examples of the resulting graphs are shown pictorially in Fig.~\ref{fig:asymmetricart}.

To avoid confusion, we note that this model is different from the well-known {\em ``directed percolation''} model (for a review see, e.g., Ref.~\cite{Hinrichsen:2000aa}).
It can also be seen as a simplified version of the dynamical model discussed above, which neglects local heat accumulation in the large $l$ limit.

 We believe that the asymmetric percolation model exhibits a percolative phase transition which falls in the usual percolation universality class for all values of $l$, as long as the critical temperatures are drawn from the distribution~(\ref{eq:PTc}), with $\alpha>0$. Therefore, this model allows us to ascribe the origin of the first order phase transition at large $\tau$ to the effect of heat accumulation.

\begin{figure}[ptb]
\includegraphics[scale=0.7]{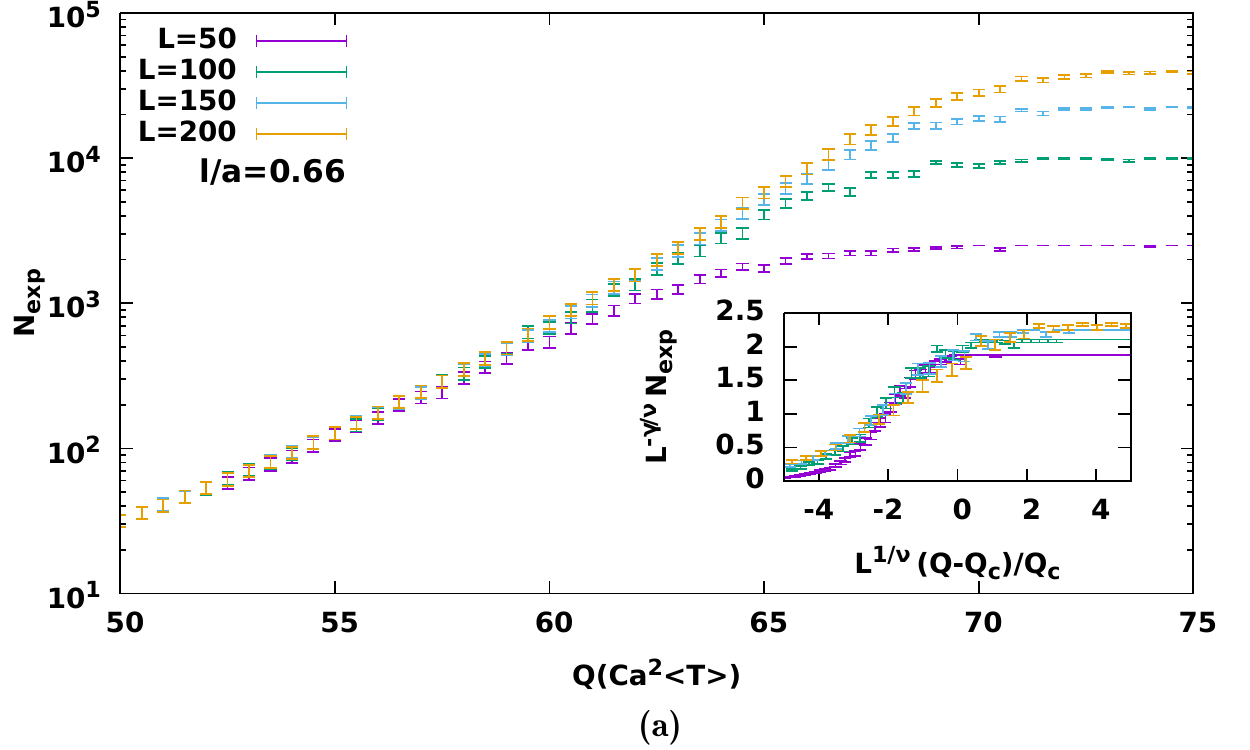}
\includegraphics[scale=0.7]{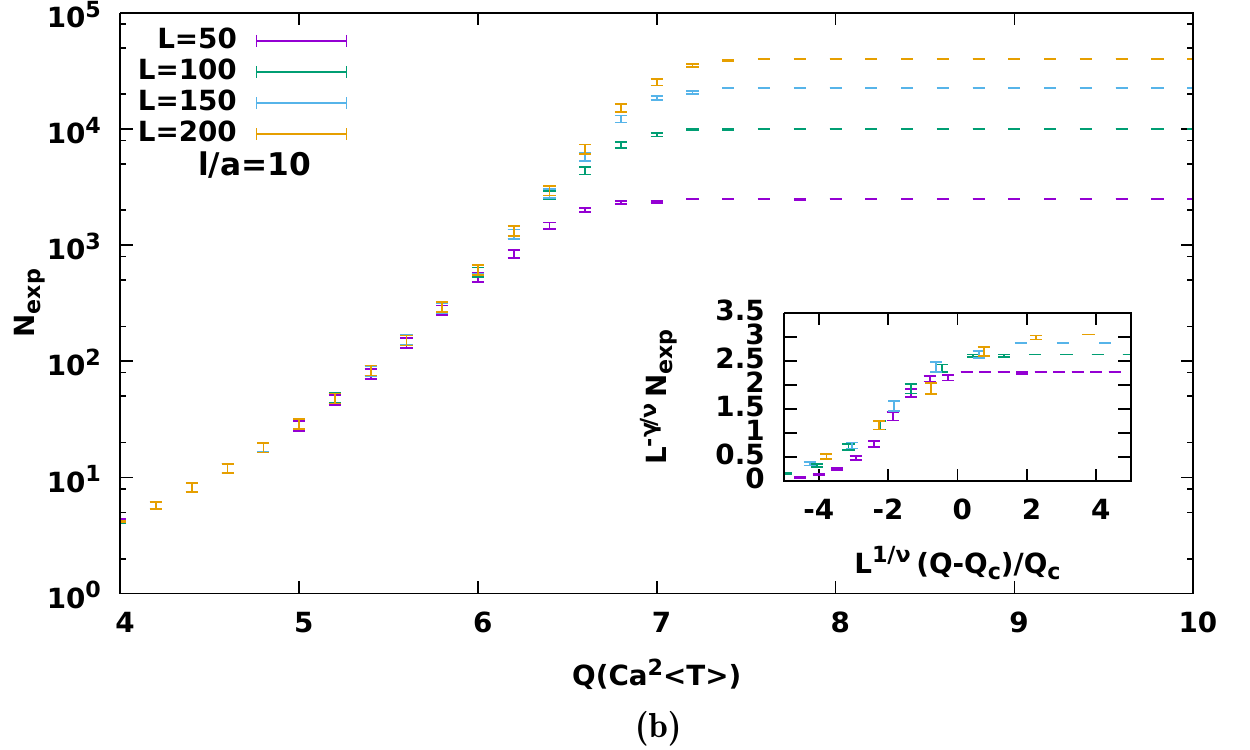}
\caption{Static asymmetric percolation model. Average size of the connected clusters, as a function of $Q$. Data are shown for (a) $\nicefrac{l}{a}=0.66$ (i.e., strong dissipation), and for (b) $\nicefrac{l}{a}=10$ (i.e., weak dissipation). The insets show the collapse (before saturation) of the data obtained via the scaling ansatz of Eq.~(\ref{eq:scaling}). In both cases, the data are compatible with a continuous transition, belonging to the same universality class as percolation.}
\label{fig:asymmetricdata}
\end{figure}

We have applied to this model the same scaling analysis described in Sec.~\ref{sec:strongdissipation}. The only difference lies in the fact that, instead of having a single exploded cluster per sample, now multiple connected clusters are present; therefore, for each sample, $\NN$ is now computed as the average cluster size. The results are shown in Fig.~\ref{fig:asymmetricdata}, where Fig.~\ref{fig:asymmetricdata}(a) shows the data for $l/a=0.66$ (i.e., strong dissipation), while Fig.~\ref{fig:asymmetricdata}(b) shows the data for $l/a=10$ (i.e., weak dissipation).

The data have been rescaled according to Eq.~(\ref{eq:scaling}). The optimal collapse is shown in the insets of the two panels. Note that, as expected,  the curves stop collapsing when the saturation value $\NN=L^2$ is reached.  We have extracted the values $Q_c/(Ca^2<T>)=70.2\pm 0.4$, $\nu=1.3\pm0.1$, and $\gamma=2.39\pm0.03$ in the case of strong dissipation. Conversely, in the weak dissipation case, we found $Q_c/(Ca^2<T>)=7.1\pm0.1$, $\nu=1.33\pm0.02$, and $\gamma=2.38\pm0.04$. The exponents are in agreement with the values predicted by percolation theory. Therefore, we conclude that the asymmetric percolation model displays no first order phase transition and is in the same universality class as percolation for all values of $l$.

\section{Conclusion}

We have shown that in the deflagration regime the system of disordered explosives has a complex phase diagram. 
Depending on the dissipation in the system, it can exhibit either a first or second order dynamical phase transition, with a tricritical point dividing the two regimes.
In the region of the second order phase transition our picture is qualitatively analogous to the one obtained for models of fire propagation (see, e.g., Ref.~\cite{Beer199077}). Indeed, such models can be viewed as the $l\rightarrow0$ limit of our model, in which heat can strictly propagate up to a finite distance.

From the perspective of disordered statistical mechanics, the first order transition is especially intriguing as quite general arguments disfavor such transitions in the presence of quenched disorder in low dimensions.
Generally, disorder smears out critical singularities as each correlation volume sees slightly different effective parameters.
Near a first order equilibrium transition, the free energy which can be gained from these fluctuations competes with the surface energy associated with the induced domain walls.
In dimensions $d \le 2$, Imry and Wortis argued that the fluctuational energy dominates and rounds any putative transition \cite{Imry:1979aa}.
These arguments were extended and made rigorous by Aizenmann and Wehr to prove that there are no first order transitions in disordered equilibrium systems in $d\le 2$ \cite{Aizenman:1989aa,Aizenman:1990aa}.
Percolation is not an equilibrium transition and these arguments do not directly apply; nonetheless, simple percolation models can be understood in terms of free energies through the Fortuin-Kasteleyn mapping onto the $q$-state Potts model, which likewise only exhibits first order behavior in high dimensions \cite{Wu:1982ra}.
First order percolative transitions have been constructed in certain long-range models with nontrivial many-body interactions in the statistical weight and/or growth process rules \cite{Achlioptas:2009aa,Riordan:2011aa,Costa:2010aa,Cho:2013aa,Maksymenko:2015aa}, but these have little to do with physical dynamics in low dimensions. A first order transition is also present in ``bootstrap percolation" models,  which are somewhat reminiscent of the heat accumulation mechanism described in this paper~\cite{ADLER1991453}.

Our results were obtained in the limit where the system is in the deflagration regime, where the speed of the front propagation is less than the speed of sound.
In this case the diffusive heat propagation can be separated from the mass motion in the medium \cite{Landau:1987ab}.
We believe, however, that our results have more general character.
First, the origin of hot spots is not necessarily described by Eq.~(\ref{heatcondEq}), which is valid only in the deflagration regime.
Several different physical mechanisms of the existence of the hot spots, such as  viscous void closure, adiabatic heating of
trapped gases, and friction, have been proposed (see, for example, Ref.~\cite{Field:1992aa}).
In particular, the hot spots can explode in the detonation regime, while the interaction between hot spots can be mediated by spherical sound or weak shock waves propagating in a ``neutral'' medium which does not detonate.
Such regimes can be considered as intermediate between deflagration and detonation.
We can call it a weak detonation regime.
The picture  presented above can be qualitatively correct in this situation as well. We also note that such a transition has been reported numerically in the detonation regime as well~\cite{TarverNichols}.  However, the authors of Ref.~\cite{TarverNichols} assume the detonation wave to be so strong to average over all local fluctuations of the system parameters: this can be seen as a mean field version of the problem addressed in this paper.

Another aspect of the problem is that,  during an explosion,   different parts of a sample can move apart and the process of the energy exchange between them terminates.
We plan to generalize our model for this case.

Finally, we would like to mention that the above results may be applicable to dynamics of any overheated disordered system with a first order phase transition.

\begin{acknowledgements}
We thank S. C. Morampudi, D. Moore, and R. Moessner for useful discussions.
 A.B. and B.S. acknowledge a financial support from Los Alamos National Laboratory, Grant No. 285149. 
C.R.L. acknowledges support from the Sloan Foundation through a Sloan Research Fellowship and the NSF through Grant No. PHY-1520535.
\end{acknowledgements}

\bibliography{explosions}

\end{document}